\documentclass[useAMS,usenatbib]{mn2e}
\usepackage[letterpaper,margin=0.65in]{geometry}
\usepackage[ansinew]{inputenc}
\usepackage{amsfonts}
\usepackage{amsmath}
\usepackage{enumerate}
\usepackage{enumitem}
\usepackage{graphicx}
\usepackage{epsfig}
\usepackage{color}
\usepackage{amssymb}

\newcommand{\eg}{{e.g., }}
\newcommand{\ie}{{i.e., }}
\newcommand{\pz}{photo-$z$\ }
\newcommand{\pzns}{photo-$z$} %pz with no trailing space, for punctuation.

\newcommand{\nit}{$n_{\rm it}$\,}

\title[SOM$z$: photo-$z$ PDFs with self organizing maps ]
{SOM$z$: photometric redshift PDFs with self organizing maps and random atlas}
\author[M. Carrasco Kind and R. J. Brunner] 
{Matias Carrasco Kind\thanks{E-mail: mcarras2@illinois.edu} and Robert J. Brunner\\
Department of  Astronomy, University of Illinois, Urbana, IL 61820 USA}

\begin{document}
\date{\today}

\pagerange{\pageref{firstpage}--\pageref{lastpage}} \pubyear{0000}

\maketitle

\label{firstpage}

\begin{abstract}
In this paper we explore the applicability of the unsupervised machine learning technique of Self Organizing Maps (SOM) to estimate galaxy photometric redshift probability density functions (PDFs). This technique takes a spectroscopic training set, and maps the photometric attributes, but not the redshifts, to a two dimensional surface by using a process of competitive learning where neurons compete to more closely resemble the training data multidimensional space. The key feature of a SOM is that it retains the \textit{topology} of the input set, revealing correlations between the attributes that are not easily identified. We test three different 2D topological mapping: rectangular, hexagonal, and spherical, by using data from the DEEP2 survey. We also explore different implementations and boundary conditions on the map and also introduce the idea of a \textit{random atlas} where a large number of different maps are created and their individual predictions are aggregated to produce a more robust photometric 
redshift PDF. We also introduced a new metric, the $I$-score, which efficiently incorporates different metrics, making it easier to compare different results (from different parameters or different photometric redshift codes). We find that by using a spherical topology mapping we obtain a better representation of the underlying multidimensional topology, which provides more accurate results that are comparable to other, state-of-the-art machine learning algorithms. Our results illustrate that unsupervised approaches have great potential for many astronomical problems, and in particular for the computation of photometric redshifts.
\end{abstract}


\section{Introduction}

Given the tremendous amount of imaging data being generated by current and forthcoming large area surveys, considerable attention has been focused on the estimation of redshifts by applying statistical techniques to  photometric observations of sources through multiple filters. To date, the Sloan Digital Sky Survey~\citep[SDSS;][]{York2000} has obtained hundreds of millions of photometric observations of extragalactic sources covering more than one quarter of the sky. This same survey has also, by using a considerably larger quantity of time with the same telescope, obtained a galaxy spectroscopic sample about one hundred times smaller albeit to a higher precision~\citep{Aihara2011}, highlighting the fact spectroscopy is both considerably more difficult and more time consuming than photometry. 

Thus, photometric redshift (hereafter \pzns s) estimation techniques provide a much higher number of galaxies with redshift estimates per unit telescope time than spectroscopic surveys~\citep{Hildebrandt2010}. Furthermore, the importance of these \pz estimation techniques will only increase with the continued development of modern, multi-band imaging surveys like the Dark Energy Survey (DES) or the Large Synoptic Survey Telescope (LSST), which probe large cosmological volumes and photometrically detect galaxies that are often too faint to be spectroscopically observed. 

Given these results, it is not surprising that the estimation of galaxy redshifts using multi band photometry has grown significantly in the last two decades. As a result, a variety of methods for computing \pzns s ~\citep[see, \eg][for a review on some current \pz techniques]{Hildebrandt2010,Abdalla2011} from template fitting techniques~\citep[see, \eg][]{Benitez2000,Ilbert2006} to empirical methods ~\citep[see, \eg][]{Collister2004,Ball2008} have been developed. In the latter category of empirical methods, the application of machine learning techniques to the estimation of \pzns s has become increasingly important in recent years. 

To date, most machine learning \pz techniques perform to a similar accuracy, while providing only a single redshift estimate and an associated error. Few public codes do exist that compute a full redshift probability density function for each galaxy~\citep[\eg][]{Cunha2009,Gerdes2010}. A \pz PDF provides more information that has been extensively shown to be important for a number of different cosmological measurements including galaxy clustering, weak lensing, baryon acoustic oscillations, and the mass function of galaxy clusters~\citep[see, \eg][]{Ho2012,Reid2010,Jee2013}. These cosmological  measurements depend strongly on both the number of galaxies in the sample and the accuracy of the measured distances to these galaxies~\citep[\eg][]{Marti2013}.
 
Given the growth of photometric surveys, these cosmological studies will require the use of reliable photometric redshifts  and a complete understanding of their uncertainties~\citep{Newman2013b}. Thus, \pz methods will be most effective going forward if they  not only provide a reliable redshift estimation but also provide a robust redshift PDF. For example,~\cite{Myers2009} have shown that by using the full \pz PDF within a two-point angular quasar correlation function, as opposed to simply using a single redshift estimate, their measurement has been improved by a factor of nearly four, which is equivalent to increasing the survey volume by a similar factor. Likewise, \cite{Mandelbaum2008} discuss how the accuracy of \pz and the inclusion of the \pz PDF affect the calibration for weak lensing studies.\\

Empirical algorithms require a spectroscopic training data set in order to train the \pz algorithm before they can be applied to new photometric observations. Some of the more recent machine learning techniques that have been applied to the \pz challenge include  artificial neural networks~\citep[\eg][]{Collister2004}, decision trees~\citep[\eg][]{Carliles2010,CarrascoKind2013a}, nearest neighbors~\citep[\eg][]{Ball2008}, Gaussian process~\citep[\eg][]{Bonfield2010}, and support vector machines~\citep[\eg][]{Wadadekar2005}. These types of algorithms also have the advantage that additional  information, such as galaxy profiles, morphology, concentration, or environmental properties, can be included in the \pz estimation process in addition to the standard galaxy magnitudes or colors. These methods, however, are only reliable within the limits of the training data, and sufficient caution must be exercised when extrapolating these algorithms beyond those limits.

All of the aforementioned techniques can be categorized as supervised learning algorithms, where the input attributes (\eg magnitudes or colors) are provided along with the desired outputs (\eg redshift), which are all employed during the learning process. In this sense, the redshift information from the training set \textit{supervises} the training phase. We recently developed TPZ\footnote{http://lcdm.astro.illinois.edu/research/TPZ.html}~\citep[Trees for Photo-Z: ][hereafter CB13]{CarrascoKind2013a}, a public \pz probability density function (PDF) estimation code that uses prediction trees and random forest techniques to compute robust \pz PDF estimations as well as ancillary information about the overall \pz estimation process for a given data set. In this supervised machine learning technique, the prediction trees use the values of the redshifts (from the spectroscopic sample) to determine the specific point and input dimension, which is an exact numerical value, at which the data will be divided into 
two branches. 
This process is repeated iteratively while building each tree in the forest.

On the other hand, an unsupervised machine learning \pz technique does not use the desired outputs (\eg redshifts from the spectroscopic sample) during the training process; thus no decisions are made based on this information. The only information used by the unsupervised algorithm are the input attributes themselves. A Self Organized Map (SOM):~\citep{Kohonen1990,Kohonen2001} is an unsupervised, neural network algorithm that is capable of projecting high-dimensional input data (\eg the dimensions might represent the magnitudes, colors or other attributes of a galaxy) onto a low-dimensional (usually two dimensions are sufficient) map ~\citep{Lawrence1999}. Thus, a SOM corresponds to a nonlinear projection of the training data that attempts to preserve the topology of the input attributes from the multidimensional space. 

Self organized maps have been utilized in several astronomical applications~\citep[\eg][]{Naim1997,Brett2004,inder2012,Fustes2013}. Recently, ~\citet{Geach2012} and ~\citet{Way2012} have introduced the application of a SOM to compute a single \pz estimator, providing strong evidence that this technique has distinct advantages and can also be extended to compute a \pz PDF. The unsupervised nature of this approach provides a complementary tool to  supervised algorithms, such as our previously published TPZ, thereby opening the possibility to develop a meta-classifier that uses multiple, complimentary approaches to improve the precision with which we can estimate \pz PDFs. Another important characteristic of a SOM when applied to \pz estimation is the ability to produce a structured ordering of the spectroscopic training data, since similar galaxies in the training sample are mapped to neighboring neural nodes in the trained feature map. The application of this technique for the classification of sources based 
on 
their location within the feature map, however, is still an underutilized tool.

In this work, we extend the previous work of~\citet{Geach2012} and ~\citet{Way2012} to use self organized maps to produce \pz PDFs and explore different configurations. This new work follows our previously published TPZ approach originally developed for decision trees and random forests. Herein, we present an ensemble learning method that generates multiple, different SOMs, and subsequently combines their results into a final output providing a probability distribution which we call SOM$z$. In analogy to the random forest technique~\citep{Breiman1996,Breiman2001}  used by TPZ, we use bootstrap samples from the training data to build a set of unsupervised independent feature maps. By aggregating the predictions from this \textit{atlas} of random maps, we produce a more accurate and robust final estimate. As introduced in CB13, our implementation incorporates the errors of the measured attributes by perturbing the galaxy parameters by their measurement uncertainties. We repeat this process, generating multiple 
individual new observations for each galaxy that are subsequently combined 
into a 
final PDF that can be used as desired to estimate a single redshift and its associated error. Furthermore, we also explore the implementation of this algorithm by using three different two-dimensional topologies: a rectangular grid, a hexagonal grid, and a spherical surface corresponding to the 2D representation of the multidimensional training sample.

This paper is organized as follows. We provide a complete and detailed  description of the SOM  method we develop and apply to compute \pzns s PDF in \S2. \S3 describes the data set we use to test the methodology of this approach, while also characterizing the efficacy and accuracy of our implementation. In \S4 we present the main results of our tests of our new approach. In \S5 we  analyze our results and discuss the capabilities of this algorithm. Finally, in \S6 we conclude with a summary of our main results while discussing some of the advantages and limitations of our new SOM$z$ algorithm.

%%%
\section{Self organized maps}

Since their introduction~\citep{Kohonen1982}, Self-Organized-Maps have been applied to a variety of scientific problems~\citep[see \eg][for a detailed description of the SOMs and some of their applications]{Kohonen2001}. A SOM is a type of an artificial neural network where the learning is unsupervised, there are no hidden layers, and a direct mapping  is produced between the training set and the output network. Another important characteristic of a SOM is that the training phase of the algorithm is a competitive process, called vector quantization,  where each node or neuron in the map competes with the other nodes or neurons to become more similar to the training data, \ie each neuron tries to represent as closely as possible the galaxy training set within each timestep. This fact and the use of a neighbor function, which modifies a region of spatially close cells, make the SOM a unique tool that preserves very closely the topology of the  multidimensional spectroscopic sample. As a result, similar nodes 
tend to 
be grouped together, where, for our purpose, each node represents galaxies with similar properties.

\begin{figure}
\includegraphics[width=0.98\linewidth]{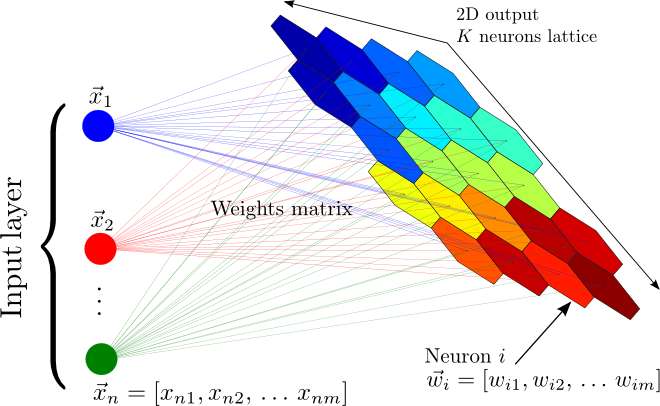}
\caption{A schematic representation of a self organized map. The training set of $n$ galaxies is mapped into a two-dimensional lattice of $K$ neurons that are represented by vectors containing the weights for each input attribute. Note that the galaxies and the weight vectors are of the same dimension $m$, and that one neuron can represent more than one training galaxy. The color of the map encodes the organization of groups of galaxies with similar properties. The main characteristic of the SOM is that it produces a nonlinear mapping from an $m$-dimensional space of attributes (\eg magnitudes) to a two-dimensional lattice of cells or neurons.} 
\label{fig:som_scheme}
\end{figure}

Figure~\ref{fig:som_scheme} presents a schematic illustration of how a SOM is trained.  During this phase, each node on the two-dimensional map can be represented by weight vectors of the same dimension as the size of the  training galaxy sample. In an iterative process, the galaxies from the training set are individually used to correct the weight vectors so that the specific neuron (or node) that, at a given moment, best represents the input galaxy is modified, along with the weight vectors of its neighboring neurons, to become a better representation of the current input galaxy. This process is repeated for every galaxy; and the SOM generally converges within a few iterations to its final form where the training data is separated into \textit{groups} of similar features, illustrated in Figure~\ref{fig:som_scheme} by colors.

There are different versions of the basic SOM algorithm; however, all of them follow the same procedure when training a map. The differences arise in the method by which the weight vectors are updated. In this paper, we present our results from testing two standard versions of the SOM algorithm mapped to three different topologies for a two-dimensional lattice.
\subsection{SOM$z$ Algorithm}\label{som_alg}
We now present a more detailed discussion of the actual SOM algorithm. First, consider a set of $n$ input vectors taken from the  galaxy training sample, which we denote by $\vec{x} \in R^{m}$. These vectors are $m$-dimensional, where each dimension is a different, measured galaxy attribute, \ie magnitudes, colors or any other information about the galaxy except the actual spectroscopic redshift. Second, consider a set of $K$ weight vectors $\vec{w_k} \in R^m$ where $k= 1,..., K$. These $K$ weight vectors, which correspond to different neurons, are arranged in a two-dimensional lattice for a given topology. Initial values for the weight vectors are drawn from a uniform random distribution.

 For every \nit iteration, all  $n$ galaxies from the training set are individually processed, and the weights are modified iteratively to optimally match each galaxy. This is the procedure which produces the self-organization of the maps and conserves the topology of the training space. When processing each 
training galaxy, the weight components of the neuron that most closely matches the current galaxy are updated, along with the weight components of the topologically closest neurons, to better represent this input entry within the featured map. The result of this direct mapping procedure is an approximation of the galaxy training probability distribution function, and it can be considered as a simplified representation of the attribute space of the galaxy sample. We have implemented two different techniques: on-line and batch, to update the actual weights of each cell.

\begin{enumerate}
\item On-line SOM: In this case, the weight vectors are updated recursively after processing each input galaxy. For each galaxy,  the Euclidean distance between the galaxy's vector of attributes (denoted by $\vec{x}$) and each neuron's weight vector from the map (denoted by  $\vec{w_k}$) is computed at a given timestep $t$:
\begin{equation}\label{dist}
 d_k(t) = d(\vec{x}(t),\vec{w_k}(t)) = \sqrt{\sum_{i=1}^m \left[x_i(t) - w_{k,i}(t)\right]^2}
\end{equation}
From this list of distances, the best matching cell, or neuron, will be identified and denoted by the subscript $b$, as the cell that is the closest  to the galaxy at timestep $t$:
\begin{equation}
 d_b(t)=\min_k d_k(t)
\end{equation}

With this technique, however, not only is the best-matching node updated but also that node's neighboring nodes. In this manner, the entire region containing the best-matching node is identified as being similar to the current training galaxy. This helps ensure similar nodes are co-located, which mimics how training galaxies that have similar properties tend to be co-located in the higher dimensional parameter space. To update the weights, we employ the following relation:
\begin{equation}\label{update1}
 \vec{w_k}(t+1) = \vec{w_k}(t) + \alpha(t) H_{b,k}(t)[\vec{x}(t) - \vec{w_k}(t)]
\end{equation}
where $\alpha (t)$ is the learning-rate factor, which is reduced monotonically for each timestep. This factor quantifies the magnitude of the correction for the cells as a function of time:
\begin{equation}\label{alpha_eq}
\alpha(t) = \alpha_s \left( \frac{\alpha_e}{\alpha_s} \right)^{t/(n_{\rm it}*n)}
\end{equation}
where $\alpha_s$ is the starting value of $\alpha$, usually close to unity,  $\alpha_e$ is the ending value, and $n_{\rm it} \times n$ is the total number of timesteps. $H_{b,k}(t)$ is the neighborhood function that also decreases with time \textit{and} with  the distance between the nodes $b$ and $k$. This function quantifies the physical extent to which nodes near to the best-matching node are also updated at every time step. The choice of the kernel's shape for this function does not significantly affect the results as the \pz PDF estimation in this iterative process. However, the kernel must be smooth, it must be symmetric to avoid biases in any direction, and it must decrease  monotonically away from the best matching node so that nodes closer to the best matching node are more strongly updated. The Gaussian Kernel is the simplest kernel that retains all of these features, therefore we use it 
 in our \pz PDF computations as:
\begin{equation}\label{H}
H_{b,k}(t) = e^{-D^2_{b,k}/\sigma(t)^2}
\end{equation}
where $D_{b,k}$ is the distance between the nodes $b$ and $k$ which depends on the topology used.  

The parameter $\sigma(t)$ encodes the width of the neighborhood function that decreases with $t$, from a value comparable to the size of the map $\sigma_0$ to roughly the width of a single cell $\sigma_f$:
\begin{equation}
 \sigma(t) = \sigma_0 \left( \frac{\sigma_f}{\sigma_0}\right)^{t/(n_{it}*n)}
\end{equation}
This procedure is applied to all $n$ training galaxies, which are processed in a random order during each iteration. This process is repeated for  \nit iterations, where just a few iterations are sufficient. As a result, the weights are updated $n_{\rm it} \times n$ times during the training process, but only the last updated weights are retained after the training process.\\

\item Batch SOM: This scheme is very similar to the on-line technique; however, in the batch method the weights are updated \textit{only} after each iteration is completed and not after each training galaxy has been processed. As a result, the order in which galaxies are processed in this approach is irrelevant. The weights $\vec{w_k} (t_{it})$ are updated at the end of each iteration for a total of \nit times by using an accumulated sum:
\begin{equation}\label{update2}
 \vec{w_k}(t_{it}) = \frac{\sum_{j=1}^{n}\widetilde{H}_{b,k}(t_{it})\vec{x_j}}{\sum_{j=1}^{n} \widetilde{H}_{b,k}(t_{it})}
\end{equation}
where the summation is over all $n$ galaxies in the training sample, and $t_{it}$ is the timestep representing a given iteration. $\widetilde{H}_{b,k}(t_{it})$ is computed by using Equation \ref{H}, but in this case the best-matching node is identified by using the weights computed at the end of the previous iteration:
\begin{equation}
\tilde{d}_k(t_{it})=d(\vec{x}(t_{it}),\vec{w_k}(t_{it-1})) = \sqrt{\sum_{i=1}^m \left[x_i(t_{it}) - w_{k,i}(t_{it-1})\right]^2}
\textbf{}\end{equation}
and
\begin{equation}
 \tilde{d}_b(t_{it})=\min_k \tilde{d}_k(t_{it})
\end{equation}
Again, recall that the weight vectors $\vec{w_k}(t_{it})$  in the batch technique are computed at the end of the previous iteration and kept fixed during the current one. In this case, the update of the weight vectors is not recursive as in the on-line technique; therefore, the final map does not depend in any way on the order in which the training galaxies are sampled. In addition, the batch technique does not use the learning-rate $\alpha(t)$, which eliminates a potential source of poor convergence if this factor is not well determined. 

\end{enumerate}

\begin{figure}
\includegraphics[width=1.1\linewidth]{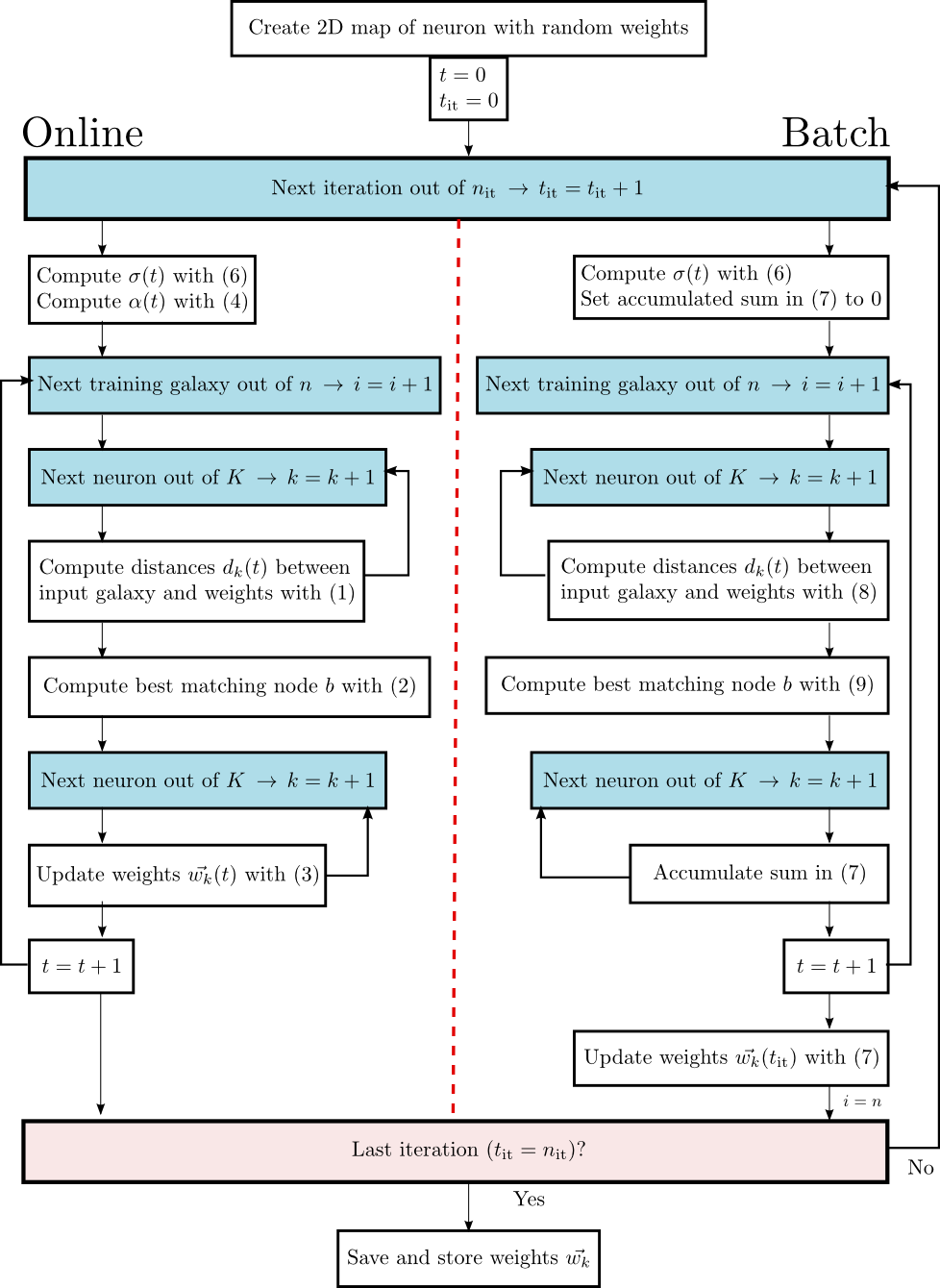}
\caption{A flowchart illustrating our implementation of the SOM algorithm for \pz estimation. Online and batch update schemes are presented on the left and right respectively.} 
\label{fig:som_diagram}
\end{figure}

Figure \ref{fig:som_diagram} illustrates the SOM algorithm and highlights the difference between the two techniques we have employed to update the weight vectors during the training process. Both techniques are initialized in the same manner, have common steps, and require similar running times. With the batch update technique, however, there is no dependency on $\alpha$ and only the neighborhood function is updated for each time step $t$.

\begin{figure*}
\includegraphics[width=0.31\textwidth,height=0.31\textwidth]{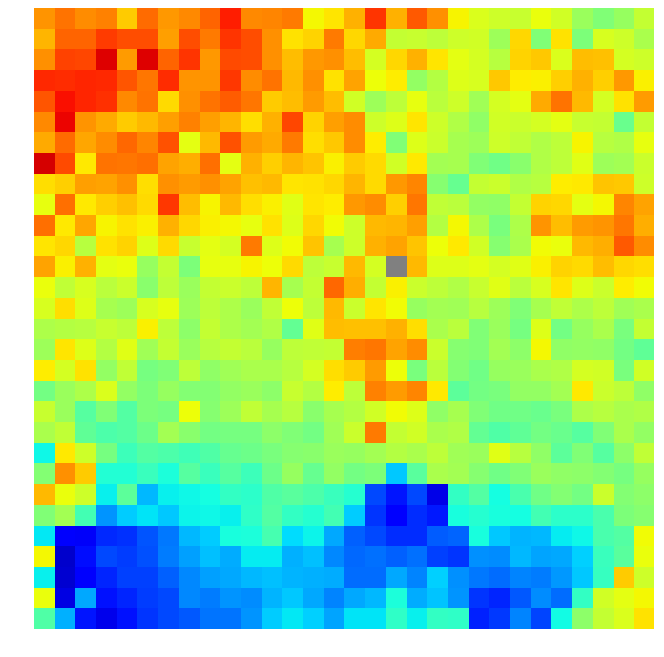}
\includegraphics[width=0.31\textwidth,height=0.31\textwidth]{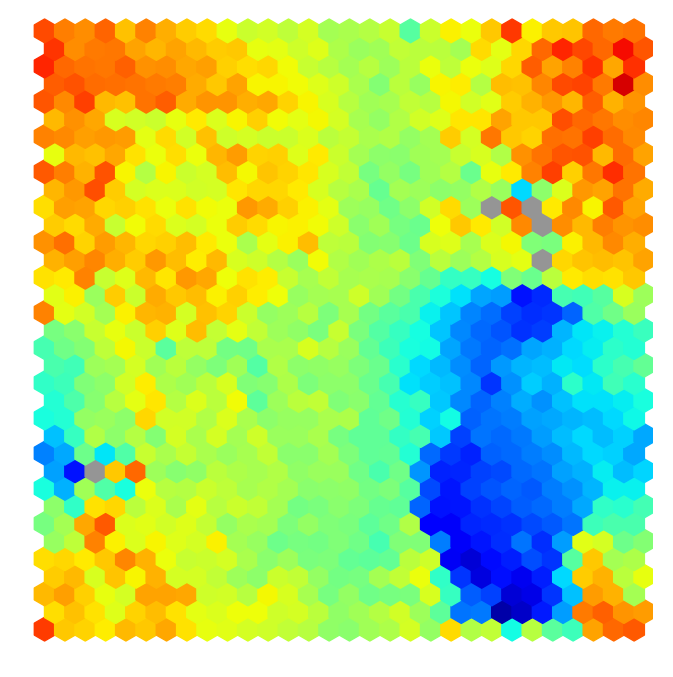}
\includegraphics[width=0.31\textwidth,height=0.31\textwidth]{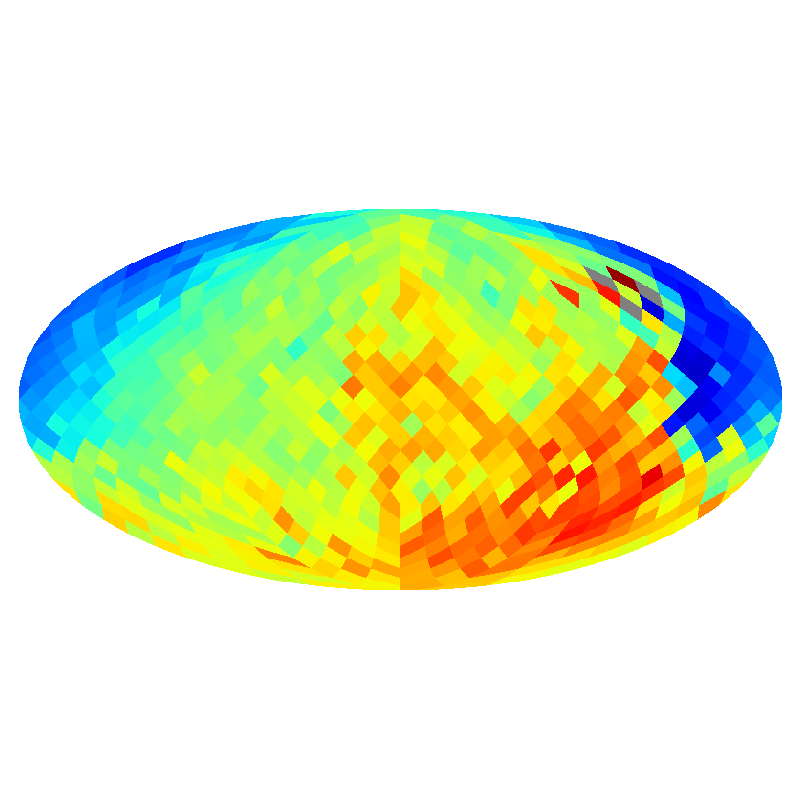}
\includegraphics[width=0.03875\textwidth,height=0.31\textwidth]{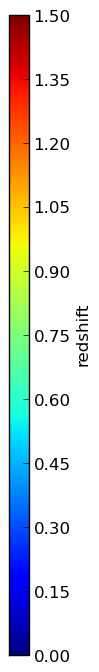}
\caption{A comparison of the three different, two-dimensional topologies used in this work. Each topology employs equal-area cells, where the color encodes the mean redshift of all galaxies assigned to a cell after the training process is complete. The colorbar on the right applies to all three maps. (\textit{Left}): Rectangular  grid with 784 square cells. (\textit{Central}): Hexagonal grid corresponding to 756 cells with periodic boundary conditions. (\textit{Right}): Spherical grid using HEALPix with 768 cells.}
\label{fig:metrics}
\end{figure*}

\subsection{2D Topologies}\label{2dtop}
For each of the two SOM techniques discussed in the previous section, we have implemented three different, two-dimensional topologies: a rectangular grid with square cells, a hexagonal grid, and a grid of equal-area cells confined to the surface of a sphere. We also include the option to use periodic boundary conditions for the non-spherical case. Figure \ref{fig:metrics} presents the nodes for these three topologies constructed via the data described in \S3. Each topology has roughly the same number of cells: 784 (rectangular), 756 (hexagonal), and 768 (spherical). For this figure we have employed the same training process using the online update scheme for each topology, and the cell colors encode the mean redshift of the galaxies represented by each cell after the last iteration has been completed. This simple visualization demonstrates how the SOM technique groups galaxies together via their input parameters, while the desired predictive attribute, in this case redshift, is only used at the end to 
visualize the 
map or to make \pz estimations. The SOM technique, for all three 
topologies, clearly groups galaxies together 
in the map that have similar redshifts without any specific supervision, which is a major advantage of this method.

We now present the details of these three, two-dimensional topologies.
\begin{enumerate}
 \item Rectangular grid: For this topology, each cell has eight direct neighbors. We calculate the distances $D_{b,k} $, which is used by $H_{b,k}(t)$, between the best-matching cell and the other cells by using the Euclidean distances $D_{b,k} = \sqrt{(x_b-x_k)^2+(y_b-y_k)^2}$.  This topology is the standard method used to create SOMs, and it has been extended by using periodic boundary conditions so the nodes are wrapped on one toroidal surface. This is functionally equivalent to folding a sheet of paper into a tube, and subsequently wrap the tube onto itself to form a torus.
 \item Hexagonal grid: For this topology, each cell has six direct neighbors. We calculate the distances $D_{b,k}$ between cells by using the Euclidean distances between the centers of the cells as in the rectangular grid topology. It differs from the rectangular grid by the fact of all the neighbor's centers are located at the same distance which produces a smoother neighboring function. This topology can also be extended by using periodic boundary conditions so that the nodes are effectively wrapped on to one surface.
\item Spherical grid: This last topology naturally eliminates the problem of wrapping the nodes as the map is constructed directly on a continuous, two-dimensional surface. For this topology, we have used the HEALPix\footnote{http://healpix.jpl.nasa.gov}~\citep{Gorski2005} tools to construct the two-dimensional map where the cells are constructed to have the same area as the other topologies. We calculate the distances between cells by using the great-circle distance between the centers of each cell:
\begin{equation}
D_{b,k} = \cos^{-1}(\sin \phi_b \sin \phi_k + \cos \phi_b \cos \phi_k \cos (|\theta_b - \theta_k|)
\end{equation}
where $\phi$ and $\theta$ are the latitude and longitude respectively of the best matching cell $b$ and the $k$ nearest cells.
\end{enumerate}

\subsection{Random Atlas}\label{RAs}

In machine learning, a random forest is an \textit{ensemble learning} algorithm that first generates many  randomized prediction trees and subsequently combines the predictions together into a meta-prediction. Random forests have been demonstrated~\citep{Caruana2008} to be one of the most accurate empirically trained learning techniques for both low and high dimensional data. Since we are using self-organized maps in this work, however, we can not construct a collection of trees as described in CB13. Instead, we explore the construction of a collection of maps, which we aggregate and call a \textit{random atlas} in a similar manner as a random forest.

Given a training sample of $n$ galaxies that have $m$ attributes (\eg magnitudes), we create $N_M$ bootstrap samples of size $n$ (\ie $n$ randomly selected objects with replacement) to generate $N_M$ different maps. For each map, we can either use all available attributes and have weight vectors of the same dimensions or, alternatively, we can randomly select a subsample of attributes for each map that reduces  possible correlations between maps. After all maps are built,  a final and robust prediction can be calculated by combining all $N_M$ estimates together. As we discussed in CB13, this technique  performs well when compared to other learning techniques and is also robust against overfitting (\ie there is no limit on the number of maps, $N_M$, in the \textit{atlas})

\subsection{SOM Implementation}
In order to generate \pz PDFs by using SOMs we have two major tasks. First, after preparing the training data, we generate $N_R$ training samples by perturbing the measured training data attributes according to the measured uncertainty for that attribute, which we assume to be normally distributed. In this manner, we can incorporate the measurement error into the map construction. We also reduce the bias towards the data and introduce randomness into the construction of the maps in a systematic manner. Second, for each newly constructed training sample, we generate $N_M$ new maps as described previously in \S\ref{RAs} by using bootstrap samples. 

In total, we produce $N_R \times N_M$ SOMs as described in \S\ref{som_alg}. After all the final weights for each map are recorded, the galaxies for each sample are processed again by using those weights and are assigned to one of the $K$ cells belonging to each map. This ensures that each cell in each map represents a subsample of galaxies that have similar characteristics. To compute a \pzns, we process each galaxy in the test sample (\ie the photometric data) and determine which cell in each map best represents this galaxy. We repeat this procedure for all SOMs; and, when this is completed, we combine the predictions from all of the maps into a single probability density function that is normalized by the total number of predictions. In this manner, each map contributes equally to the final PDF. 

\begin{figure*}
\includegraphics[width=0.94\textwidth]{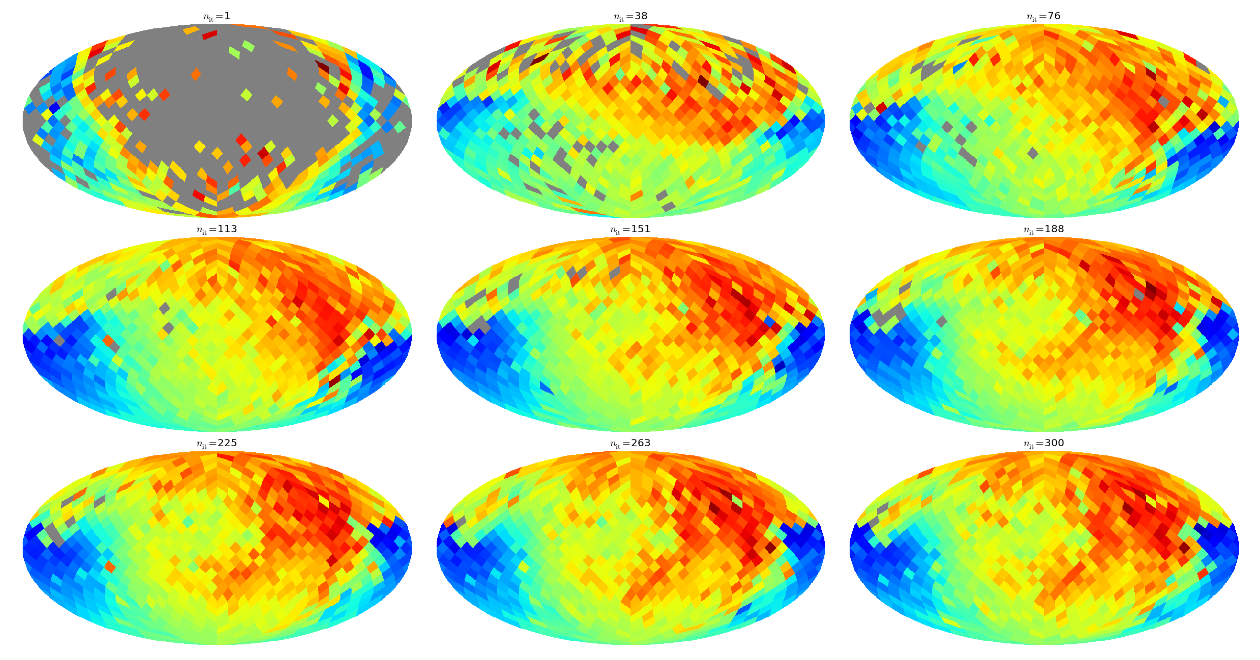}
\raisebox{15ex}{\includegraphics[width=0.03875\textwidth,height=0.31\textwidth]{Figures/colorbar.png}}
\caption{Evolution of the SOM at different iterations using the spherical topology and online updating. Colors encode the mean redshift of the galaxy being represented by each cell at each iteration as defined by the colorbar, similar to  the one in Figure~\ref{fig:metrics}.}
\label{fig:evol}
\end{figure*}

This process is demonstrated for one example map in Figure~\ref{fig:evol}, where the evolution of one SOM is sampled at different iterations using online updating. As before, the colors encode the mean redshift of the galaxies represented by each cell during each iteration. From this figure, we see that even at the first iteration there is a slight separation that quickly changes with time until convergence to the final distribution is achieved and galaxies with similar redshifts are spatially grouped in a self-organized manner.

\section{DATA}\label{deep2data}
To explore different configurations and to demonstrate the capabilities and the efficacy of this SOM \pz approach, we follow the example we presented in CB13, but in this paper we restrict our analysis to data from the Deep Extragalactic Evolutionary Probe (DEEP) survey.  

The DEEP survey is a multi-phase, deep spectroscopic survey that was performed with the Keck telescope. Phase I used the Low Resolution Imaging Spectrometer (LIRS) instrument~\citep{Oke1995}, while phase II used the DEep Imaging Multi-Object Spectrograph (DEIMOS)~\citep{Faber2003}. The
DEEP2 Galaxy Redshift Survey is a magnitude limited spectroscopic survey of objects with $R_{AB} < 24.1$~\citep{Davis2003,Newman2013a}. The survey includes photometry in three bands from the Canada-France-Hawaii Telescope (CFHT) 12K: $B$, $R$, and $I$ and it has been recently extended by cross-matching the data to other photometry databases. In this work, we use the Data Release 4~\citep{Matthews2013}, the latest DEEP2 release that includes secure and accurate spectroscopy for over 38,000 sources. The photometry for the sources in this catalog was expanded by using two $u$, $g$, $r$, $i$, and $z$ surveys: the Canada-France-Hawaii Legacy Survey~\citep[CFHTLS;][]{Gwyn2012}, and the SDSS. For additional details about the photometric extension of the DEEP2 catalog, see~\cite{Matthews2013}. 

To use the DEEP2 data with our SOM implementation, we have selected sources with secure redshifts (ZQUALITY$\geq 3$), which were securely classified as galaxies, have no bad flags, and have full photometry. Even though the filter responses are similar, the $u$, $g$, $r$, $i$, and $z$ photometry come from two different surveys and are thus not identical. We therefore treat those galaxies with SDSS photometry for fields 2, 3, and 4 of the DEEP2 target areas independently from those for field 1 with CFHTLS photometry. In the end, this leaves us with a total of 20,227 galaxies with eight band photometry and redshifts, from this data set randomly select 10,000 for training and hold the rest for testing.

\section{Results}\label{results}

In this section, we compare the results of our SOM implementation by using different parameter configurations with the DEEP2 data.  To compare different applications of this algorithm, we define the bias to be $\Delta z' = |z_{\rm phot}-z_{\rm spec}|/(1+z_{\rm spec})$, and we present the standard metrics used to compare the accuracy of the different SOMs in Table \ref{tab:def_metrics}. As shown in this table, we define several metrics to address the bias and the variance of the results (the first five rows) and also present three values to characterize the outlier fraction. 

We have introduced the quantity $KS$, which represents the results of a Kolmogorov--Smirnov test to address whether the predicted \pz distribution and the spectroscopic redshift distribution are drawn from the same underlying population. We present this new statistic since it provides one robust value to compare both distributions that does not depend on how we bin in redshift and it is defined as the maximum distance between both empirical distributions. For this statistic, we compute the empirical cumulative distribution function (ECDF) for both distributions. For the spectroscopic sample the ECDF is defined as:
\begin{equation}
F_{\rm spec} (z) =  \sum\limits_{i=1}^{N} \Omega_{z_{\rm spec}^i < z}
\end{equation}
where N is the number of galaxies in the redshift sample, and
\begin{equation}
 \Omega_{z_{\rm spec}^i < z} = \begin{cases} 1, & \mbox{if } z_{{\rm spec},i} < z  \\ 0 , & \mbox{otherwise } \end{cases}
\end{equation}
The summation is carried out over all galaxies in the sample. Having computed the ECDF for both the \pz and spectroscopic distributions, we compute the KS statistic as:
\begin{equation}
{\rm KS} = \max_z \left(\lvert\lvert F_{\rm phot} (z) - F_{\rm spec} (z)\rvert\rvert \right)
\end{equation}
As a result, as the KS statistic decreases, the two distributions become more similar.
\begin{table}
\caption{Definition of the metrics used in the text to present and discuss the results, }
\label{tab:def_metrics}
\centering
\renewcommand{\footnoterule}{}
\begin{tabular}{ll}
Metric & Meaning\\
\hline
$<\Delta z'>$ & mean of $\Delta z'$\\
$|\Delta z'|_{50}$ & median of $\Delta z'$ \\
$\sigma_{\Delta z'}$ & Standard deviation of $\Delta z'$ \\
$\sigma_{68}$ & Sigma value at which 68\% of $\Delta z'$ is enclosed \\
$\sigma_{\rm MAD}$ & Median absolute deviation = ${\rm median}(||\Delta z' - |\Delta z'|_{50}||)$\\
${\rm KS}$ & Kolmogorov - Smirnov statistic for $N(z)$\\
${\rm out}_{0.1}$ & Fraction of outliers where $\Delta z' > 0.1$\\
${\rm out}_{2\sigma}$ & Fraction of outliers where $|\Delta z' - <\Delta z'> | > 2\sigma_{\Delta z'}$\\
${\rm out}_{3\sigma}$ & Fraction of outliers where $|\Delta z' - <\Delta z'> | > 3\sigma_{\Delta z'}$\\

\end{tabular}
\end{table}

All of the metrics listed in Table \ref{tab:def_metrics} are defined such that a lower value for the computed metric indicates a better overall \pz solution. We have defined a new, meta-statistic, which we call $I$-score (symbolically represented by $I_{\Delta z'}$), to more easily compare different SOM parameter configurations (\ie online or batch and a specific 2D topology) or different \pz estimation techniques. For this new meta-statistic, we first must normalize each set of metrics across all different \pz estimations so that we are not biased by different dynamic ranges. Thus, for example, we first compute the mean and standard deviation for  $<\Delta z'>$, and subsequently rescale all individual $<\Delta z'>$ values so that this set of values has zero mean and unit variance. 

We continue this process for all nine statistics listed in Table~\ref{tab:def_metrics}, and compute their weighted sum to obtain the $I$-score: 
\begin{equation}
I_{\Delta z'} = \sum \frac{M_i}{w_i}, 
\end{equation}
where $M_i$ is the rescaled metric and weight value for metric $i$ out of the nine available. For simplicity, we use equal weights in the remainder of this paper (and thus the $I$-score is simply the average of the nine rescaled metrics for each technique). As a result, the \pz method (or parameter configuration) with the lowest $I$-score will be the optimal estimation technique. On the other hand,  if  we are looking for a technique or parameters configuration with, for instance, a lower outlier fraction, we could assign a higher weights accordingly to account for it. In this way, we can efficiently select the best method or configuration for specific needs.

\section{Discussion}
In order to explore the effects of different parameter configurations on the performance of our SOM \pz implementation, we conducted twenty different tests and compare their $I$-score results in Table~\ref{tab:big_metrics} by using six colors from the DEEP2 data: $B-R$, $R-I$, $u-g$, $g-r$, $r-i$, and $i-z$. These configurations include the use of all three topologies discussed in \S\ref{2dtop}: Hexagonal (hex), Rectangular (rec) and Spherical (sph); the use of \textit{online} or \textit{batch} methods to update the weights as shown in Figure \ref{fig:som_diagram}; and the use of a Random Atlas where different maps are built using random subsets (random = yes) of four colors or single maps where all six colors are used (random = no), which gives twelve different configurations. In addition, both rectangular and hexagonal topologies were used with both periodic and non-periodic boundary conditions (spherical is by definition wrapped), which gives us an additional eight configurations.

We determined the best values for the other parameters in our SOM \pz implementation, which were then fixed for all twenty tests, by using an Out-Of-Bag data (similar to a validation sample) technique we presented in CB13. For example, we set the values $\alpha_s$ and $\alpha_e$  in Equation \ref{alpha_eq} to be 0.9 and 0.5 respectively. In addition, each random atlas contains 100 different maps and each topology contained approximately 800 cells in a given map. All galaxies in the test sample were used for each run. The results, averaged over ten realizations, are presented in Table \ref{tab:big_metrics} for all the metrics, where we have used the mean redshift in place of each PDF for simplicity. Note that the last column is the $I$-score. For clarity, we highlight in red the best value for a particular metric.

\begin{figure}
\includegraphics[width=0.98\linewidth]{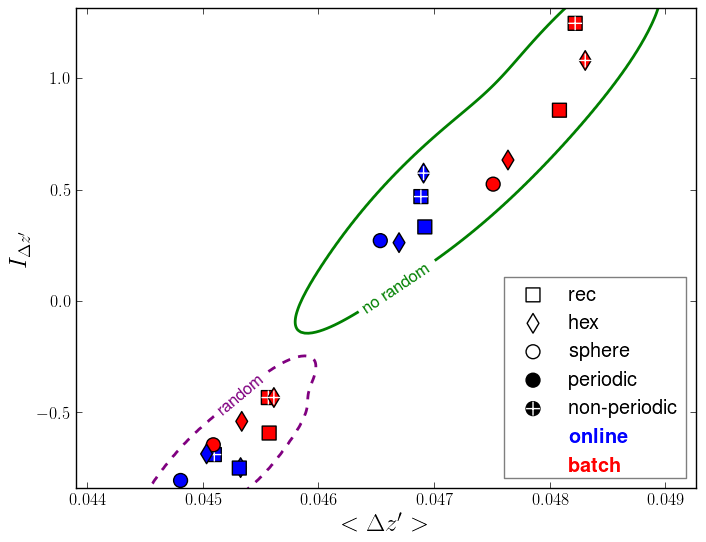}
\caption{The $I$-score, $I_{\Delta z'}$ as a function of the bias, $<\Delta z'>$ for all twenty methods discussed in the text,  averaged over ten different realizations for all galaxies (10,227) in the test sample. Enclosed by a green curve are the results of using all attributes on each map of the atlas and enclosed in purple the results when a random atlas was used. Blue symbols indicate an \textit{online} update of the weights, while red symbols indicates a \textit{batch} update. The symbols themselves represent different topologies and the white cross indicates periodic boundary conditions. } 
\label{fig:I_z_bias}
\end{figure}

We compare these different parameter configurations visually in Figure \ref{fig:I_z_bias}, where the twenty runs are plotted in terms of their bias and $I$-score values. In this figure, different symbols represent different topologies (squares for rectangular, diamonds for hexagonal and circles for spherical), colors represent the update method used (blue for online update and red for batch update). 

The curves enclose all test results that either use a random subsample of attributes (purple) or all attributes (green) on each map inside the atlas. Note that the separation of these two groups of tests is a direct output from our SOM \pz implementation. Finally, we highlight if periodic boundary conditions were used for rectangular or hexagonal topology by a white cross.

Overall, from both Table \ref{tab:big_metrics} and Figure~\ref{fig:I_z_bias}, the best set parameter configuration is spherical topology with an \textit{online} update using random atlas. This run has metrics that are the closest to the best values and it has the lowest $I$-score value. In the rest of this section, we explore the results of these different parameter configurations in more detail.

\begin{table*}
\begin{minipage}[]{\textwidth}
\caption{Results table for all the twenty combinations averaged after ten different realizations. The red entries show the best value on each column to aid the reading of the table.}
\label{tab:big_metrics}
\centering
\renewcommand{\footnoterule}{}
\begin{tabular}{l l l l c c c c c c c c c c}
\hline
Topology & Periodic\footnote{P stands for periodic boundary topology and NP for non-periodic} & update & random & $<\Delta z'>$ & $|\Delta z'|_{50}$ & $\sigma_{\Delta z'}$ & $\sigma_{68}$ & $\sigma_{\rm MAD}$ & ${\rm KS}$ & ${\rm out}_{0.1}$ & ${\rm out}_{2\sigma}$& ${\rm out}_{3\sigma}$ & $I_{\Delta z'}$\\
 \hline  \hline 
rec & NP & online & no & 0.0469 &0.0280 &0.0685 &0.0353 &0.0194 &0.0635 &0.1042 &0.0322 &0.0152 &0.4733 \\
 \hline 
rec & NP & online & yes & 0.0451 &0.0273 &0.0667 &0.0338 &0.0188 &0.0719 &0.0956 &0.0298 &0.0147 &-0.6868 \\
 \hline 
rec & NP & batch & no & 0.0482 &0.0291 &0.0709 &0.0358 &0.0202 &0.0636 &0.1075 &0.0321 &0.0150 &1.2525 \\
 \hline 
rec & NP & batch & yes & 0.0456 &0.0277 &0.0673 &0.0340 &0.0190 &0.0742 &0.0970 &0.0295 &0.0145 &-0.4305 \\
 \hline 
hex & NP & online & no & 0.0469 &0.0281 &0.0685 &0.0353 &0.0195 &0.0628 &0.1045 &0.0325 &0.0153 &0.5789 \\
 \hline 
hex & NP & online & yes & 0.0453 &0.0275 &0.0674 &0.0339 &0.0189 &0.0728 &0.0962 &{\color{red}0.0289}& 0.0141 &-0.7454 \\
 \hline 
hex & NP & batch & no & 0.0483 &0.0290 &0.0717 &0.0359 &0.0200 &0.0630 &0.1078 &0.0311 &0.0147 &1.0855 \\
 \hline 
hex & NP & batch & yes & 0.0456 &0.0278 &0.0674 &0.0340 &0.0191 &0.0734 &0.0970 &0.0292 &0.0145 &-0.4302 \\
 \hline 
sph & P & online & no & 0.0465 &0.0277 &0.0685 &0.0351 &0.0193 &0.0626 &0.1031 &0.0324 &0.0150 &0.2753 \\
 \hline 
sph & P & online & yes & {\color{red}0.0448}& 0.0272 &0.0669 &{\color{red}0.0337}& {\color{red}0.0187}& 0.0718 &0.0961 &0.0296 &0.0143 &{\color{red}-0.8034} \\
 \hline 
sph & P & batch & no & 0.0475 &0.0287 &0.0696 &0.0356 &0.0198 &0.0625 &0.1057 &0.0311 &0.0143 &0.5293 \\
 \hline 
sph & P & batch & yes & 0.0451 &0.0274 &{\color{red}0.0664}& 0.0338 &0.0189 &0.0732 &0.0970 &0.0300 &0.0144 &-0.6428 \\
 \hline 
rec & P & online & no & 0.0469 &0.0281 &0.0695 &0.0352 &0.0195 &0.0651 &0.1041 &0.0314 &0.0145 &0.3370 \\
 \hline 
rec & P & online & yes & 0.0453 &0.0273 &0.0674 &0.0338 &0.0188 &0.0738 &0.0962 &0.0292 &0.0141 &-0.7474 \\
 \hline 
rec & P & batch & no & 0.0481 &0.0293 &0.0708 &0.0358 &0.0201 &0.0655 &0.1075 &0.0308 &0.0140 &0.8618 \\
 \hline 
rec & P & batch & yes & 0.0456 &0.0277 &0.0673 &0.0339 &0.0191 &0.0745 &0.0975 &0.0291 &{\color{red}0.0139}& -0.5900 \\
 \hline 
hex & P & online & no & 0.0467 &0.0279 &0.0691 &0.0351 &0.0194 &{\color{red}0.0615}& 0.1038 &0.0319 &0.0148 &0.2663 \\
 \hline 
hex & P & online & yes & 0.0450 &{\color{red}0.0272}& 0.0670 &0.0337 &0.0189 &0.0725 &{\color{red}0.0948}& 0.0296 &0.0147 &-0.6838 \\
 \hline 
hex & P & batch & no & 0.0476 &0.0289 &0.0704 &0.0357 &0.0199 &0.0619 &0.1070 &0.0310 &0.0141 &0.6378 \\
 \hline 
hex & P & batch & yes & 0.0453 &0.0275 &0.0672 &0.0339 &0.0190 &0.0739 &0.0970 &0.0296 &0.0144 &-0.5375 \\
 \hline 
\end{tabular}
\end{minipage}
\end{table*}

\subsection{Random atlas }

The first parameter configuration we examine is the use of a random atlas. As shown in Table \ref{tab:big_metrics} or Figure \ref{fig:I_z_bias}, there is a clear, albeit numerically small difference between the performance of our SOM algorithm with and without the use of random subsampling of attributes. This finding is remarkably similar to the result we discussed in CB13 where a random forest was shown to be superior to prediction trees that used the full set of attributes. The likely explanation is the random sampling of attributes when building the maps (trees) for a random atlas (forest) more completely explores the set of attribute combinations than when using all attributes. 

These maps are constructed using Bootstrap sampling; thus by definition all maps are different although they are likely to be highly correlated, which will yield stable results after a certain number of maps have been generated. When using random sampling of the attributes, however, we are by definition introducing extra variation into the algorithm. This can reduce the noise variables that will always contribute when all attributes are included, and will on average yield better statistics when a large number of maps are generated so that all variables are used multiple times in different combinations for different maps. For example, if we construct 100 maps where each map is constructed by randomly selecting four color attributes out of six possible colors, we can be sure all attribute combinations (in this case 15) are sufficiently covered.

The appropriate number of attributes to be randomly selected when constructing a random atlas can be determined either by testing the algorithm using Out-Of-Bag~\citep{CarrascoKind2013a} data on previous runs or by selecting a value somewhere between the total number of attributes and the number of dimensions of the SOM (in this case we have a two-dimensional topology). Alternatively, in CB13 we discussed using $\sqrt{M}$, where $M$ is the set of attributes, although this is likely too small for lower dimensional problems. A reasonable compromise might be to simply use $2/3$ of the attributes when constructing each map.

We test the dependence of the random atlas on the number of input attributes by constructing two hundred maps using a spherical topology with online updating and only changing the number of attributes that are used for the random sampling. The results are presented in Table~\ref{tab:atts}, where our $I$-score statistic indicates that four attributes are optimal (as well as two other metrics). However, it is interesting to note that three attributes perform only moderately worse than four, and that two attributes show comparable performance to either five or six attributes. We found similar results by using other parameter configurations (\ie varying the topology and update method), suggesting the optimal number of attributes is dependent on the data themselves.

One last observation from the data presented in Table~\ref{tab:big_metrics} is that all metrics have their lowest values when using random sampling, expect for the KS statistic. This means that, on average, using all attributes produces an $N(z)$ from the training sample that seems to be a better match to the spectroscopic sample than when using random subsamples (\ie the  $N(z)$ ECDFs are more similar). This is most likely a result of the fact that a random atlas prediction produces a \pz PDF that has a smaller bias and scatter (as shown in Table~\ref{tab:big_metrics}) and is thus more strongly peaked about the mean value than a \pz PDF that does not use random sampling. When simply using the mean value from a PDF, the  $N(z)$ ECDF will thus be more strongly concentrated about the mean leading to a higher $KS$ statistic. As we will show, by using the full \pz PDF when constructing the sample $N(z)$, we generate a more realistic redshift distribution that reduces the $KS$ statistic by a 
factor of a few, reinforcing this interpretation.

\begin{table}
\caption{Performance of the SOM algorithm by using a spherical topology with an online update for different number of attributes used in the construction of the random atlas}
\label{tab:atts}
\centering
\renewcommand{\footnoterule}{}
\begin{tabular}[h!]{crrrrr}
 \hline
 Attributes & $<\Delta z'>$ & $\sigma_{\Delta z'}$ &  ${\rm KS}$ & ${\rm out}_{0.1}$ &  $I_{\Delta z'}$\\
 \hline \hline 
1 &0.0903 &0.1023 &0.1852 &0.3153 &1.6830 \\
 \hline 
2 &0.0558 &0.0659 &0.1229 &0.1379 &-0.2452 \\
 \hline 
3 &0.0446 &{\color{red}0.0607 }&0.0846 &0.0885 &-0.4636\\
 \hline 
4 &{\color{red}0.0432 }&0.0610 &0.0767 &{\color{red}0.0784 }&{\color{red}-0.4754 } \\
 \hline 
5 &0.0422 &0.0633 &0.0614 &0.0824 &-0.2726 \\
 \hline 
6 &0.0436 &0.0653 &{\color{red}0.0583 }&0.0886 &-0.2262 \\
 \hline 
\end{tabular}
\end{table}

\subsection{Weights updating}
The second parameter we explore is the method used to update the weights that control how cells in the final topology are modified. Both Table \ref{tab:big_metrics} and Figure \ref{fig:I_z_bias} demonstrate that the online weight updating method consistently performs better, when all other parameters are kept fixed, than the batch updating method. We interpret this difference as a manifestation of the dynamic nature of online updating, where the weights are updated after analyzing each galaxy, as opposed to the once an iteration update that is performed with the batch method. As a result, the online method will easily converge given a sufficient number of iterations and will produce a more accurate topological mapping. On the other hand, the batch method is nearly parameter free, while  the online method depends on the parameter $\alpha$. In addition, the batch method is computationally faster since the number of weight updates is considerably smaller than the online method, and the batch method is easier 
to scale to big data since the processing is inherently parallel.

\subsection{Topologies}

The next parameter we tested is the type of two-dimensional topology used for the SOM. Although not as obvious as the previous two parameters, the results suggest that spherical topology is superior than either the rectangular or the hexagonal grids, when given approximately equal number of cells and when using a random atlas (we note that when using the full attribute set the hexagonal topology slightly outperforms the spherical topology). Given the nature of the spherical topology, a direct comparison is only realistic when we compare to periodic boundary conditions for rectangular and hexagonal topologies. 

Although we explore the effect of the map size on the performance of this algorithm in subsection~\ref{s-sizes}, we note that since we use the HEALPIX scheme to characterize the spherical topology, we are implicitly constrained in the number of cells in our final map, which is given by $n_{\rm cells} = 12 \times {\rm nside}^2 $ where ${\rm nside}$ is a power of 2. For these tests, we used ${\rm nside}$ = 8 which corresponds to 768 cells. By using this relation, the next map size for a spherical topology would be 3072 cells, which nearly equals the number of galaxies in our training sample, and is, therefore, too large for this particular problem. This is a limitation of the spherical topology, as both the rectangular and hexagonal topologies are not restricted in this manner; thus we might be able to fine tune the number of cells in these alternative topologies in order to outperform the spherical topology. We do not, however, test this hypothesis in this paper.

Since we use the HEALPIX representation for spherical topology, we consider that the natural periodicity of this topology is superior to the forced periodicity with the rectangular and hexagonal topologies. HEALPIX generates equal-area cells and the cell centers are naturally aligned along the same latitude. Thus, it is reasonable to expect that the HEALPIX scheme produces cell weights  that more closely match the spectroscopic data set. On the other hand, we have no natural driver to choose between rectangular or hexagonal, and, depending on the value of the other algorithm parameters one may outperform the other.

\subsection{Periodicity}

Next, we look at the use of periodic boundary conditions, which from the results presented in Figure \ref{fig:I_z_bias} and Table \ref{tab:big_metrics} appear to, in general, outperform the non-periodic boundary conditions (with the understanding that the spherical topology is implicitly periodic).  Specifically,  when comparing periodic (solid colors) and non periodic cases (white crosses) in Figure~\ref{fig:I_z_bias} for the hexagonal and rectangular topologies, the periodic case performs slightly better, although this is not universally true. 

While the redshift distribution of our training sample data are limited to the range $0 \leq z \leq 1.5$, the SOM mapping process has no such restrictions when optimizing the topological mapping, for example when processing the colors of the galaxies. On the other hand, non-periodic conditions might work better for classification problems where clear separation is desired between classes of objects (\eg star versus galaxy); but in a regression problem, like \pz estimation, a clear separation is not necessarily desired as we do not want to bias the mapping either away from or specifically towards any particular region of the parameter space.

\subsection{Other parameters}

Besides the previously discussed parameters, our SOM algorithm does not depend on many other parameters (and of course the SOM can be applied in a non-parametric manner). One parameter that must be specified when using online updating, however, is the learning rate factor, $\alpha$. This parameter quantifies the correction applied to each cell at each time step, and can take values from 1.0 (maximal correction) to some minimum value, often close to 0.5. In the end, the SOM algorithm is not extremely sensitive to this parameter as the neighborhood function exerts more control over the corrections applied to the neighboring cells. We do acknowledge that, if the number of iterations is limited, this factor might become more important since fewer corrections will be applied.

Another parameter to specify is the number of iterations to use when constructing the SOM, as we need a sufficient number to generate a map that truly represents the data appropriately. This number will depend on both the size of the input data set and the number of cells in the map. For the example discussed herein, we found that 100 iterations were sufficient. For a larger data problem, the number of iterations should be increased, with the exact value determined empirically by, for example, terminating the iterative process if the map changes by some value (\eg $1\%$) or if the map evaluation does not change beyond some small tolerance.

As an example, Figure~\ref{fig:evol} demonstrates how a spherical topology map changes in nine different steps during 300 iterations using the online updating scheme. In each map, the color of a cell encodes the mean redshift of all galaxies within that cell after each evaluation. After the first evaluation, the map is not fully populated, thus only some of the cells are populated. The iteration process, however, quickly begins to populate the cells and by iteration 113 (middle left image) the map becomes fairly stable with only a few empty cells. The last three maps (iterations 225, 263, and 300 left-to-right in the bottom row) are nearly identical, demonstrating how the iterative process has essentially converged and the map can be used for \pz predictions.

%%%%%%%

\subsection{Size of map and size of atlas}\label{s-sizes}

\begin{figure}
\includegraphics[width=0.98\linewidth]{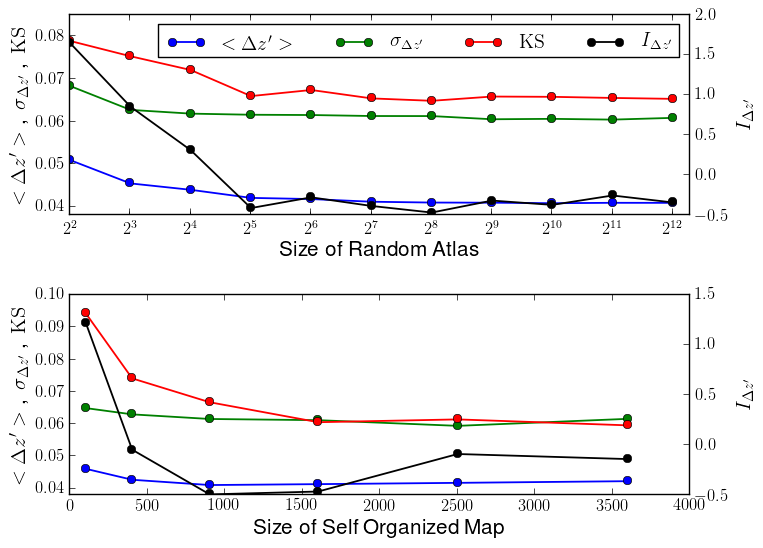}
\caption{Bias $<\Delta z'>$ (blue), scatter $\sigma_{\Delta z'}$ (green), $KS$ (red) and $I_{\Delta z'}$ (black) as a function of the number of maps contained in the random atlas with fixed map size of 756 cells (top panel), and as a function of map size keeping fixed the number of maps (100) in the random atlas (bottom).} 
\label{fig:sizes}
\end{figure}

The two  algorithm parameters that remain to be identified are the number of cells to use within an individual map, and the number of maps to use within a random atlas. We explore the effects of changing one of these parameters while keeping the other fixed in Figure \ref{fig:sizes}, where the top panel identifies the dependence on the number of maps used in a random atlas, keeping the size of each map fixed at 756 cells, while the bottom panel highlights the dependence  on the number of cells in an individual map, keeping the number of maps fixed at 100. In both panels, we use four metrics to quantify the performance of the SOM:  the bias $<\Delta z'>$ (blue), the scatter $\sigma_{\Delta z'}$ (green), the $KS$ statistic (red), and the $I$-score $I_{\Delta z'}$  (black).

%%% Copied from topology section
%%%

As shown in the top panel, where we have constructed a SOM using hexagonal topology with online updating and a fixed number of cells in each map, increasing the number of maps in the random atlas does improve the performance and reduces the value of these metrics. At some point, however, adding more maps does not produce any improvements as all possible parameter combinations have been included in the atlas a few times  and new maps become redundant. On the other hand, as shown in the bottom panel, increasing the number of cells with a single map will also improve the value of the metrics.

\begin{figure}
\includegraphics[width=0.98\linewidth]{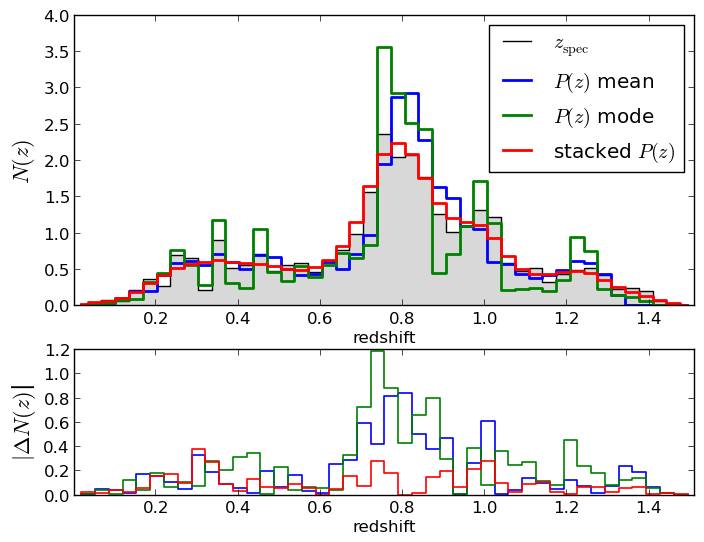}
\caption{$N(z)$ (top) and absolute error (bottom) for the galaxies used to compute Figure \ref{fig:true}, showing the difference by using the mean, the mode and a stacked \pz PDF} 
\label{fig:Nz}
\end{figure}

\begin{figure*}
\includegraphics[width=0.48\linewidth]{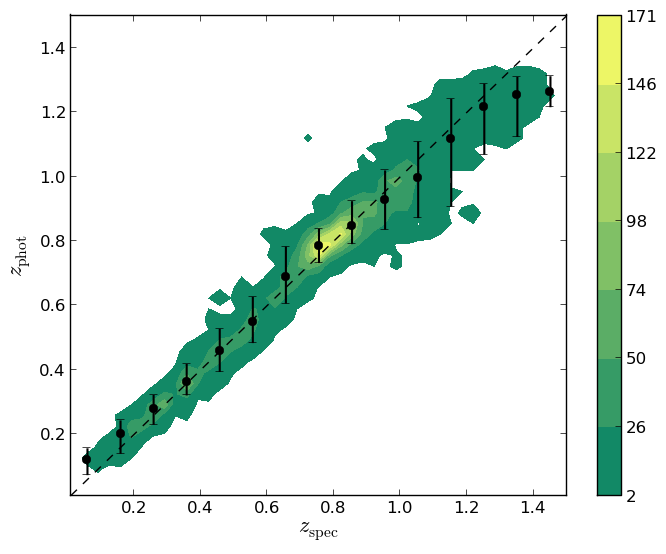}
\includegraphics[width=0.48\linewidth]{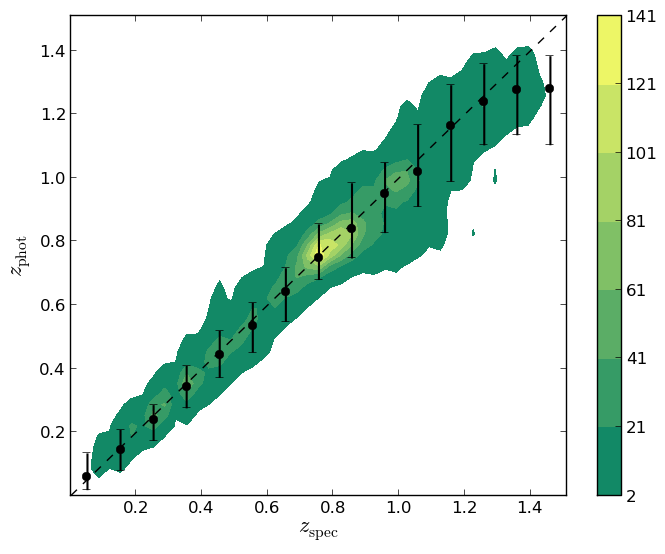}
\caption{Spectroscopic redshift versus photometric redshift estimated by using a SOM for (left) the mean of the \pz PDF and (right) the full \pz PDF. In both panels, we use an identical number of pixels to construct the image and also use the same number of contours to present the color-mapping. The galaxies used to make this image were selected in an identical manner with $zConf > 0.7$ (with a total of 8387 galaxies) from the DEEP2 survey. The black dots are the median values of $z_{\rm phot}$ and the errors bars correspond to the tenth and ninetieth percentiles within a given spectroscopic bin of width $\Delta z = 0.1$.} 
\label{fig:true}
\end{figure*}

Eventually, however, the mean number of sources per cell decreases to the point where we have empty cells with no predictive power, and we also suffer from over-fitting. This primarily affects the fraction of outliers and subsequently the $I$-score, which depends on all of the metrics. Thus we find an optimal size (for this particular data set) of approximately 1500 cells. This confirms  the results presented by~\citet{
Way2012} who also found that 
increasing the number of cells in a map produces better results until over-fitting affects the metrics.

\subsection{Photo-$z$ PDF using SOMs}

For simplicity, to this point we have compared the different SOM configurations and overall performance by simply using a single predictive value (in this case, the mean value of the \pz PDF). As we discussed in \S\ref{som_alg}, however, the SOM technique generates a full probability distribution function for the photometric redshift of each individual galaxy. We can use all of the information encoded in these PDFs when making cosmological measurements. For example, we can more accurately compute the sample redshift distribution, $N(z)$, which is used in a variety of cosmological measurements, by including the full \pz PDF. 

This can be seen from the data in Table \ref{tab:big_metrics}, where the best $KS$ statistic, which is a measurement of how well the true $N(z)$ is recovered, is 0.0615, which was computed by using the mean of the PDF. If on the other hand we use the full \pz PDF, this same metric value is 0.0221, which is almost a factor of three better. This value of a $KS$ statistic is traditionally interpreted in that we cannot reject the null hypothesis (that both the spectroscopic and the photometric distributions are the same) at a 5\% level. 

We explicitly compare the spectroscopic $N(z)$ to the measured $N(z)$ distributions computed by using the mean of the \pz PDF (blue line), the mode of the \pz PDF (green line), and the full \pz PDF (red line) in the top panel of Figure \ref{fig:Nz}.  In addition, the bottom panel displays the absolute error between the spectroscopic $N(z)$ and these three different measured $N(z)$ distributions. In both panels, the full \pz PDF is clearly shown to more closely match the spectroscopic distribution, a result that we also saw in CB13 with \pz PDFs generated by using TPZ. This simple test highlights the power of using the full information provided by a \pz PDF.

In general,  computing other metrics, such as the bias or scatter, by using the \pz PDF will produce slightly larger values than simply using the mean of the \pz PDF, since for these simpler metrics the mean is a sufficient estimator of the full PDF, while the full PDF adds information from other bins, decreasing the precision to which these metrics are computed. While these metrics are primarily useful in merely characterizing the approximate accuracy of the algorithm, it is still important that these metrics are symmetric and unbiased as a function of redshift (which confirms the lack of any systematic biases in the algorithm). 

Following our previous definition of $zConf$ from CB13 (\ie the integrated probability between $z_{\rm phot} \pm \sigma_{SOM}(1+z_{\rm phot})$, where we have set the expected scatter $\sigma_{SOM} = 0.075$), we compare spectroscopic versus photometric redshift in Figure~\ref{fig:true} by using the mean of the PDF (left panel) and full PDF (right panel) for exact same 8387 galaxies selected to have $zConf > 0.7$. Both panels are constructed from the same galaxy sample, share the same number of pixels, and have the same number of contours (although the dynamic range of the contours varies). The over plotted black dots and error bars convey the median and tenth and ninetieth percentiles, respectively, for spectroscopic bins of width $\Delta z = 0.1$

By construction, the galaxies used for Figure~\ref{fig:true} all have a concentrated \pz PDF, and as shown in Figure~\ref{fig:true}, by using the mean of the \pz PDF we have a tight, symmetric relationship. This reaffirms the conclusion found in CB13 ---but this time for a SOM \pz PDF--- that the $zConf$ value can be used to identify galaxies with accurate \pz estimates. Although the \pz PDF provides a more accurate $N(z)$ relationship, by using the mean of the \pz PDF we generate a slightly tighter correlation. On the other hand, the full \pz PDF generally produces a more symmetric distribution, which can be seen both from this figure and the median values, except for the last two bins that suffer from low numbers as seen in Figure \ref{fig:Nz}. As a result, the final choice of using the full PDF or a particular statistics characterizing the full \pz PDF should be empirically quantified as it will likely depend on the particular problem under study.

\subsection{Comparison with TPZ}

As SOM$z$ used herein is an unsupervised learning method, it can be illustrative to compare this new method to an existing, supervised learning method. As we borrow many techniques in this paper from our random forest technique outlined in CB13, in this section we compare the performance of SOM$z$ with TPZ, specifically focusing on the results computed by using both methods for the DEEP2 dataset compiled by \cite{Matthews2013}. The SOM results were produced by using a random atlas with spherical topology and online updating, while the TPZ results were produced by using the regression mode with 100 trees and $m_{*} = 3$ (explained in CB13) generating PDFs of the same redshift resolution $\Delta z = 0.012$. As the SOM results were generated by using galaxy colors, we ran TPZ by using the same colors and the same training set used to generate the SOM results. 

We present a summary of key statistics from these two estimation methods in Table~\ref{tab:TPZ_SOM}. From the end results of each technique, we create three subsamples by splitting on the $zConf$ value. The first observation from these values is the similar performance of both techniques, which is somewhat surprising given the differences between the two algorithms. On the other hand, it seems likely that the randomness inherent in our implementation of the random forest and the random atlas algorithms both improve the performances of these algorithms to a similar degree. When constructing a \pz PDF, the full multi-dimensional space is subdivided (TPZ is a supervised process while SOM$z$ is an unsupervised process) into smaller volumes, which, in both cases, contain galaxies with similar properties, that are subsequently used to make redshift predictions.

The second observation from this table is that for some metrics the random forest implementation is superior, while for others the random atlas implementation wins. Given this observation, and the inherent differences between the two approaches, it seems reasonable to want to explore the combination of the predictions from disparate learning methods. We have already started to address this issue by exploring how the performance of the photo-$z$ approach is improved when combining techniques~\citep{CarrascoKind2013b}. We defer further discussion of this topic to future paper (Carrasco Kind \& Brunner, in preparation.), where we will explore the development of meta-classifiers that combine supervised, unsupervised, and template-fitting techniques to make more accurate \pz PDF estimations.

\begin{table}
\caption{A summary of key metrics that were computed by using the same datasets for solutions provided by the SOM algorithm and TPZ. The number in parenthesis is the $zConf$ value used on each case.}
\label{tab:TPZ_SOM}
\centering
\renewcommand{\footnoterule}{}
\begin{tabular}[h!]{crrrrr}
 \hline
 Method & $<\Delta z'>$ & $\sigma_{\Delta z'}$ &  ${\rm KS}$ & ${\rm KS}_{PDF}$ &  ${\rm out}_{0.1}$\\
 \hline \hline 
SOM (0.5)  & 0.0417  & 0.0608 & 0.0659 & 0.0311 & 0.0803\\
 \hline
TPZ (0.5)  & 0.0408  & 0.0640 & 0.0352 & 0.0175 & 0.0808 \\
\hline
SOM (0.7)  & 0.0382  & 0.0586 & 0.0621 & 0.0307 & 0.0660 \\
 \hline 
TPZ (0.7)  & 0.0374  & 0.0594 & 0.0320 & 0.0162 & 0.0664 \\
 \hline 
SOM(0.9)   & 0.0318  & 0.0520 & 0.0620 & 0.0304 & 0.0427\\
 \hline 
TPZ (0.9)  & 0.0306  & 0.0516 & 0.0294 & 0.0157 & 0.0430\\
 \hline  
\end{tabular}
\end{table}

\section{Conclusions}\label{sec_conc}

While the computation of a redshift by using a galaxy's magnitudes or colors has been performed in a variety of different approaches, improvements are still possible. In this paper, we have presented a new approach that computes a \pz PDF with similar performance as other machine learning techniques that we call SOM$z$. However, this new approach is an unsupervised machine learning algorithm that uses Self-Organized-Maps, which project the muti-dimensional space of attributes (magnitudes or colors) to a 2 dimensional map that attempts to conserve the topology of the higher dimensional data. Each neuron or cell in the map is updated after each galaxy is processed by means of weights that are iteratively corrected in order to better represent the training data. The spectroscopic target information is not used at all in the process of building the maps, although it is used to identify the galaxies that belong to a cell in order to make predictions from the two-dimensional map. While SOMs have been used 
previously 
to make a \pz 
prediction~\citep{Geach2012,Way2012}, in this paper we introduce the concept of a \textit{random atlas}, in analogy to a random forest for decision trees, in which a number of different maps are created 
whose individual predictions are subsequently aggregated to produce a final \pz PDF.

In this paper, we explore the different configurations that can be used to build a SOM, and introduce a new metric, the $I$-score, which efficiently takes into account different metrics indicators of the overall  performance, such as, the bias, the scatter or how well the photometric redshift distribution matches the spectroscopic one, in order to differentiate these configurations. We found that by using a random subsample of attributes to build different maps we can produce a significantly better solution than by using all the attributes, which works similarly to the random forest for prediction trees. We explored two approaches to updating the weights for each cell in the final two-dimensional map. The first technique is called online updating, in which all weights are updated dynamically after processing each galaxy. The second is called batch updating, in which the galaxy weights are applied \textit{en masse} after each iteration. Our testing indicates that online updating produces more accurate \pz 
estimates, but is also harder to parallelize, which can be a limitation when applied to very large data sets. On the other hand, the batch method is easier to parallelize as the cumulative work can be done in blocks on different cores, but is slightly less precise since the weights are updated less frequently.

The SOM process constructs a two-dimensional topology; thus we also explored three different representations for this final map: rectangular, hexagonal, and spherical grids. While the rectangular and hexagonal grids showed similar performance, we found that the spherical grid performed slightly better, likely due to its natural lack of boundary conditions that avoid biases near the edge of the grid. For the other two, flat grids we also imposed wrapped, periodic boundary conditions, but they still perform slightly worse than the spherical topology when using approximately the same number of cells. Overall, however, we do see that wrapping the two-dimensional grids (either naturally, or with imposed periodic boundary conditions) provides a better two-dimensional representation of the multi-color space occupied by the galaxies in our analysis.

On the other hand, other SOM parameters had less of an effect on the final \pz calculation. First, the number of iterations must be large enough to allow convergence. Second, if using the online method, the degree of correctiveness needs to be close to unity, while the batch method lacks this parameter. We also explored the effect of the number of cells used to construct a given map and the number of maps in a random atlas. For the former, we find that an improvement in the \pz estimation can be achieved, but the process is limited, depending on the training data volume, as eventually we can suffer from over--fitting. For the latter, we found that after a few hundred maps, the random atlas has effectively been populated by all possible parameter configurations and no new information is added with additional maps. For our demonstration example, we found an ideal combination of approximately 1500 cells and 200 maps produced the optimal \pz predictions.

As we discussed previously in CB13, by using TPZ, which employs random forests and decision trees in a supervised learning process, we found that the mean of the \pz PDF provides a good estimation of an individual \pzns. On the other hand, for a number of cosmological measurements, such as the galaxy redshift distribution, $N(z)$, better results are obtained by using all of the information encoded in the \pz PDFs. Overall, the distribution of most metrics are symmetric and the galaxy \pz distribution more accurately resembles the spectroscopic redshift distribution when using the full \pz PDF. Overall, the full \pz PDF contains more information that can be used for a variety of measurements, it also allows us to more accurately estimate the \pz error and to define likelihood cuts such as our $zConf$ parameter that selects only those galaxies with concentrated \pz PDFs. By doing so, we have shown that we can create a sample of galaxies that are less affected by catastrophic outliers.

While our new, unsupervised approach presented herein performs to a similar degree of accuracy as previous, supervised techniques, including our own TPZ algorithm presented in CB13, there are different strengths and weaknesses of each approach. As a result, we have begun to explore how to optimally combine different \pz estimation techniques, including the use of supervised, unsupervised, and template fitting techniques into a meta-algorithm to both produce more accurate \pz estimates as well as an improved identification of prediction outliers. This work, which will be presented in a future paper, is interesting as we will be able to capitalize on the strengths of these different, yet complimentary approaches while minimizing their individual weakness.

\section*{Acknowledgements}
The authors thank the referee for a careful reading of the manuscript and for comments that improved this work.
RJB and MCK acknowledge support from the National Science Foundation  Grant No. AST-1313415. MCK has been supported by the Computational Science and Engineering (CSE) fellowship at the University of Illinois at Urbana-Champaign. RJB has been supported in part by the Institute for Advanced Computing Applications and Technologies faculty fellowship at the University of Illinois.

The authors gratefully acknowledge the use of the parallel computing resource provided by the Computational Science and Engineering Program at the University of Illinois.  The CSE computing resource, provided as part of the Taub cluster, is devoted to high performance computing in engineering and science. This work also used resources from the Extreme Science and Engineering Discovery Environment (XSEDE), which is supported by National Science Foundation grant number OCI-1053575.

Funding for the DEEP2 Galaxy Redshift Survey has been
provided by NSF grants AST-95-09298, AST-0071048, AST-0507428, and AST-0507483 as well as NASA LTSA 
grant NNG04GC89G.

\bibliographystyle{mn2e}
\bibliography{SOM}

\bsp
\label{lastpage}
\end{document}